\documentclass[letter]{aa}
\usepackage{txfonts,textcomp}
\usepackage{graphicx}
\usepackage{color}
\usepackage{tikz}
\usepackage{amssymb}
\usetikzlibrary{shapes,arrows}
\usepackage{multirow}
\usepackage{natbib}
\usepackage{pdflscape}
\bibpunct{(}{)}{;}{a}{}{,}

\begin{document}

\title{Asteroseismic versus Gaia distances: A first comparison}

\author{
J.~De~Ridder\inst{1}
\and G.~Molenberghs\inst{2,3}
\and L.~Eyer\inst{4} 
\and C.~Aerts\inst{1,5}
} 

\institute{
Instituut voor Sterrenkunde, KU Leuven, Celestijnenlaan 200D, B-3001 Leuven, Belgium
\and
I-BioStat, Universiteit Hasselt, Martelarenlaan 42, B-3500 Hasselt, Belgium
\and
I-BioStat, KU Leuven, Kapucijnenvoer 35, B-3000 Leuven, Belgium
\and 
Observatoire de Gen\`eve, Universit\'e de Gen\`eve, 51 Ch.~des Maillettes, CH-1290 Versoix, Switzerland
\and
Department of Astrophysics, IMAPP, Radboud University Nijmegen, PO Box 9010, 6500 GL Nijmegen, The Netherlands
}

\authorrunning{J.~De Ridder et al.}

\titlerunning{Asteroseismic versus Gaia distances}

\date{Received / Accepted}

\abstract
{
The {\it Kepler\/} space mission led to a large number of high-precision time series
of solar-like oscillators. Using a Bayesian analysis that combines asteroseismic techniques
and additional ground-based observations, the mass, radius,
luminosity, and distance of these stars can be estimated with good precision. This has given a
new impetus to the research field of galactic archeology.
}
{
The first data release of the Gaia space mission contains the TGAS (Tycho-Gaia Astrometric Solution) catalogue with parallax
estimates for more than 2 million stars, including many of the {\it Kepler\/} targets. Our goal is to make a first proper comparison of asteroseismic and astrometric parallaxes of a selection of dwarfs, subgiants, and red giants observed by {\it Kepler\/} for which asteroseismic distances were published. 
}
{
We compare asteroseismic and astrometric distances of solar-like pulsators using
an appropriate statistical errors-in-variables model 
on a linear and on a logarithmic scale. 
}
{
For a sample of 22 dwarf and subgiant solar-like oscillators, the TGAS
parallaxes considerably improved on the Hipparcos data, yet 
the excellent agreement between
asteroseismic and astrometric distances still holds.
For a sample of 938 {\it Kepler\/} pulsating red giants, the TGAS parallaxes are
much more uncertain than the asteroseismic ones, making it worthwhile to
validate the former with the latter. From errors-in-variables modelling we find a significant discrepancy between the TGAS parallaxes and the asteroseismic values.
}
{ 
For the sample of dwarfs and subgiants, the comparison between astrometric and
asteroseismic parallaxes does not require a revision of the stellar models on
the basis of TGAS. For the
sample of red giants, we identify possible causes of the discrepancy, which we will
likely be able to resolve with the more precise Gaia parallaxes in the upcoming releases.
}


\maketitle

\section{Introduction}

The seismic study of stars has undergone a revolution during the past decade, thanks to
the space missions CoRoT \citep[launched in 2006;][]{Auvergne2009} and {\it
  Kepler\/} \citep[launched in 2009;][]{Borucki2010}. Not only did these space
data confirm the method of asteroseismology \citep[for an extensive monograph;
see][]{Aerts2010}, they also allowed powerful applications to thousands of stars
across stellar evolution for a wide variety of stellar birth masses. Major
breakthroughs of relevance to the current study of stellar distances were the
discovery of acoustic non-radial pulsation modes in red giants
\citep{DeRidder2009} and the excitation of dipole mixed modes probing both the
deep interior and the structure of the outer envelope of such stars
\citep[e.g.][]{Beck2011}.  Thanks to their mixed gravity and acoustic
character, mixed modes allow  the core properties of a star to be tuned and therefore
can be used to pinpoint the evolutionary status \citep{Bedding2011}.

Asteroseismology of red giants offers the unique opportunity of providing stellar
ages for studies of the Milky Way, termed {galactic archeology\/}
\citep[e.g.][]{Miglio2013}.  Indeed, the measurement of the frequency at
maximum oscillation power and of the large frequency separation, along with a
spectroscopic estimate of the effective temperature, can be transformed into
high-precision estimates of the stellar mass and radius by assuming that the
input physics of solar models is also applicable to solar-like stars.
Under this
reasonable assumption, stellar masses and radii can be derived with relative
precisions of merely a few per cent, while further comparison with stellar models
provides a seismic age estimate with a precision below 20\% when
systematic uncertainties due to modest variations in the input physics are taken
into account \citep{Chaplin2014,Metcalfe2014}.  Proper computation of the
apparent CoRoT or {\it Kepler\/} magnitude according to the passbands of these satellites
 then allows the luminosity of the stars to be transformed into an
``asteroseismic'' distance \citep{SilvaAguirre2012,Rodrigues2014,Anders2016}.

So far, the asteroseismic distances of stars in the solar neighbourhood have
been compared {a posteriori\/} with Hipparcos values for the distances
whenever available, with good agreement \citep[e.g.][]{SilvaAguirre2012}. With
the Gaia mission in full swing, we foresee a quantum leap forward in this
research, both in the number of targets and  the precision in measuring  the distance. After
five years of nominal monitoring, the Gaia distance estimates are expected to be
so precise that they can serve {as input\/} to improve the physics of
stellar interiors, leading to model-independent radii and better ages than
currently available as input for
exoplanet studies and galactic archeology. Here we take a first step to compare
asteroseismic distances with the astrometric values by considering the first Gaia
data release  \citep[Gaia DR1; e.g.][]{Brown2016, Prusti2016, Lindegren2016}.

\section{Nearby {\it Kepler\/} dwarfs and subgiants}

In their original study to verify asteroseismically determined parameters with
Hipparcos distances in a self-consistent way, \citet{SilvaAguirre2012}
investigated 22 dwarfs and subgiants having Hipparcos parallaxes with a relative
error better than 20\%.  These stars are close neighbours of the Sun, with
distances ranging from 20 to 260 pc, and are not known to have companions.  
In their seismic modelling method,
\citet{SilvaAguirre2012} included corrections for reddening in an iterative
approach based on distance-dependent integrated maps of extinction. The seismic
distance estimates are also mildly dependent on the adopted metallicity and this
was taken into account in their estimates of the uncertainty for the seismic distance.

\begin{figure}
    \centering
    \includegraphics[width=0.5\textwidth]{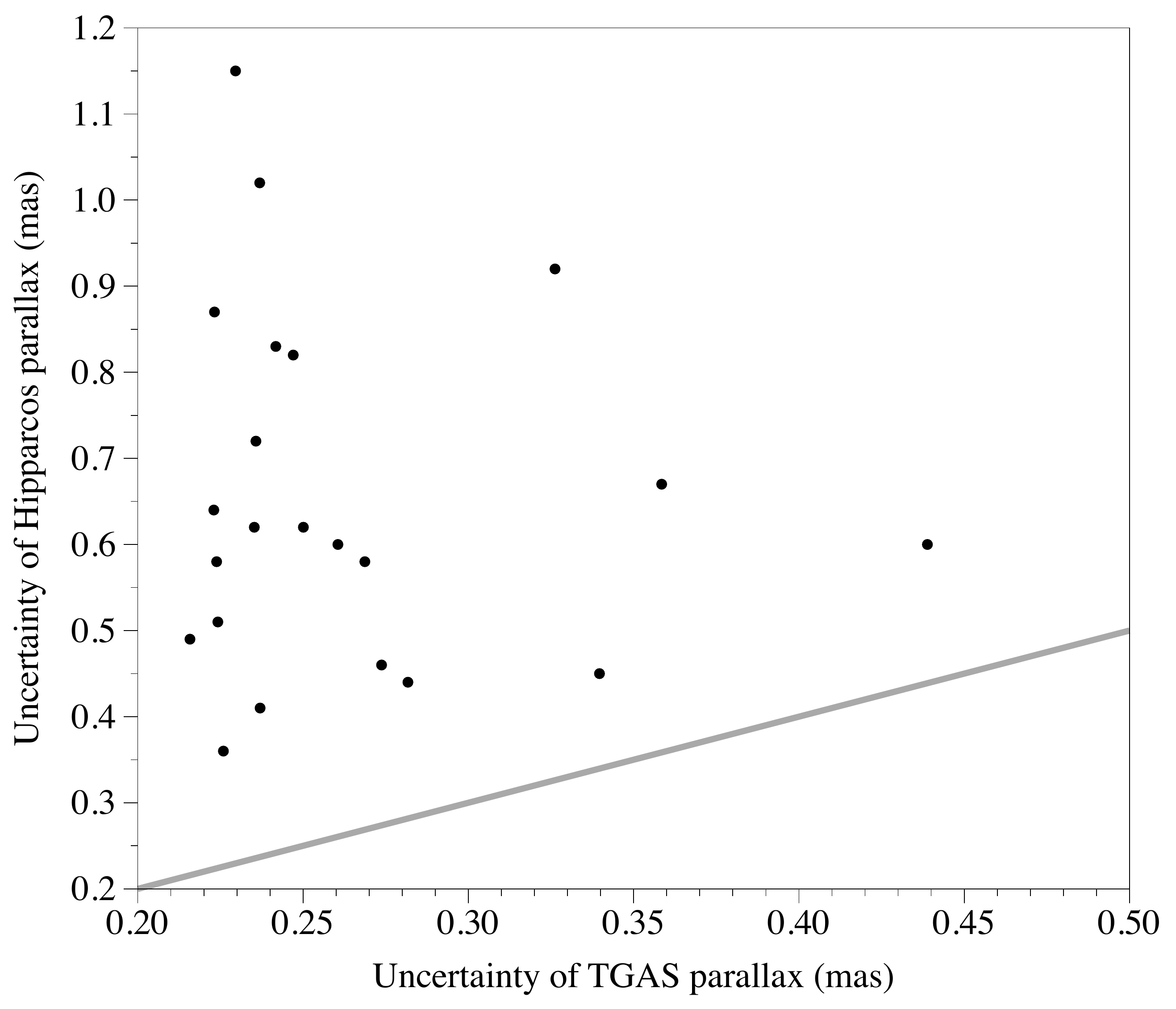}
    \caption{Uncertainty of the TGAS and Hipparcos parallaxes of the sample of
      22 nearby pulsators. The grey line is the bisector.}
    \label{fig:HipVsTGASuncertainties}
\end{figure}

\begin{figure}
    \centering
    \includegraphics[width=0.5\textwidth]{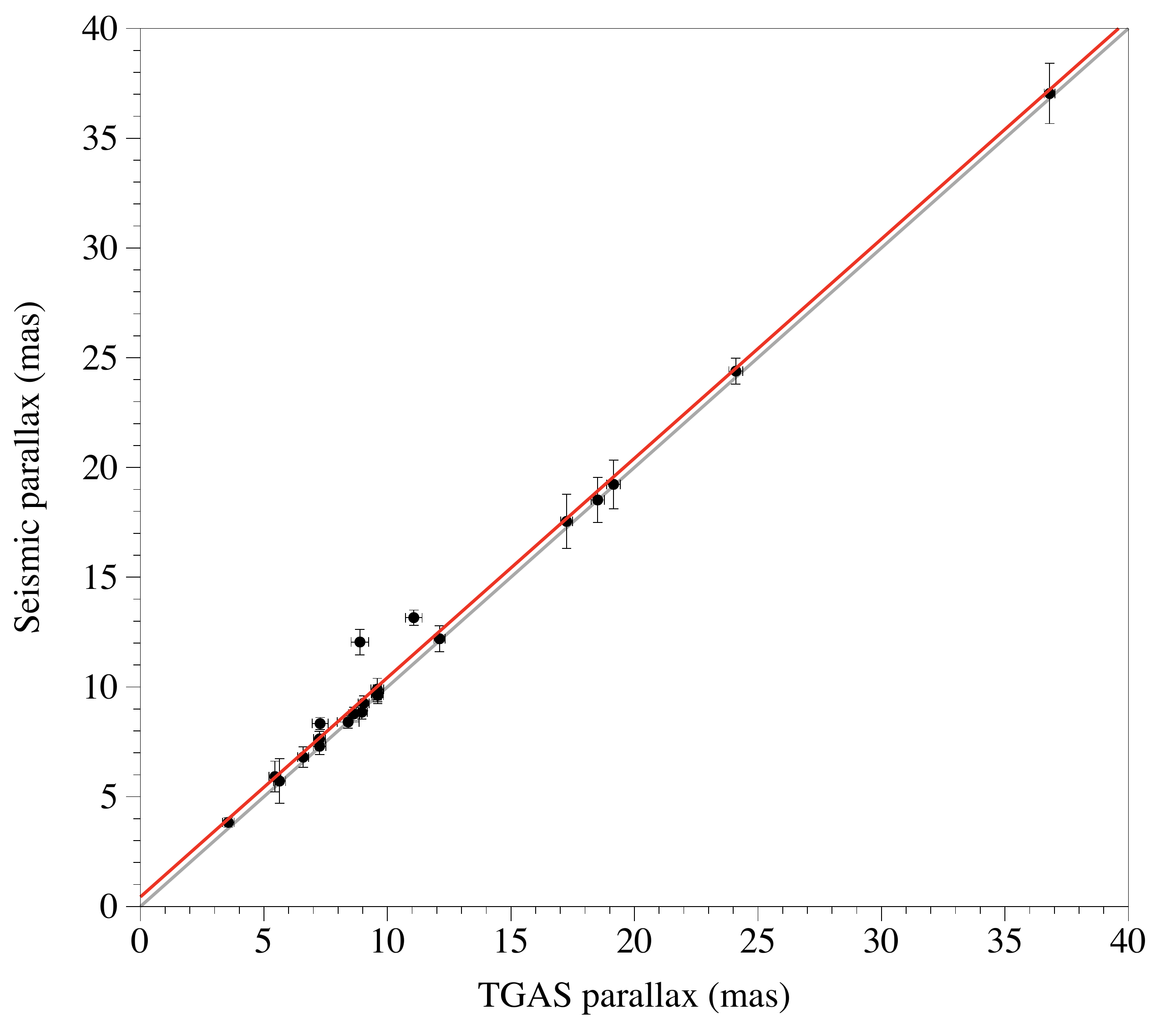}
    \caption{TGAS vs seismic parallaxes of the sample of 22 nearby pulsators. The
      grey line is the bisector. The red line is the fit using the errors-in-variables
      model (\ref{modelObservedQuantities}) with uncertainties in both measures
      taken into account.}
    \label{fig:TGASvsSeismoPar}
\end{figure}

Figure~\ref{fig:HipVsTGASuncertainties} shows the uncertainties of the parallaxes
obtained with TGAS versus those corresponding with Hipparcos.
The TGAS parallaxes considerably
improve the Hipparcos values, making a new comparison between astrometric and
asteroseismic parallaxes appropriate. In Fig.~\ref{fig:TGASvsSeismoPar} we show
the comparison between the two.
The overall correspondence between the two parallax determinations is
excellent, showing the remarkable achievement of asteroseismology. 
The best fit, shown in red, is slightly offset with respect to~the bisector, but as we  show in the following, the offset is not nearly significant enough to justify
a revision of the seismic models.

The statistical problem at hand is one where we wish to verify a linear relationship between the true parallaxes $\varpi_i^\ast$ obtained from TGAS and the true parallaxes $S_i^\ast$ derived using seismology:
\begin{equation}
\label{linearRelation}
S_i^\ast = \alpha+\beta \varpi_i^\ast .
\end{equation}
Neither $\varpi_i^\ast$ nor $S_i^\ast$ is observed. Instead, the corresponding observed quantities $\varpi_i$ and $S_i$ can be
modelled with 
\begin{align}
\varpi_i &= \varpi_i^\ast+\eta_i,    \label{twee} \\
S_i      &=  S_i^\ast+\varepsilon_i, \label{drie}   
\end{align}
where $\eta_i$ and $\varepsilon_i$ are the measurement errors. Such a model is
called an errors-in-variables model (also known as the~measurement error model) and has
been studied extensively in the statistics literature, see e.g.~\citet{WAF87}.
Here, we have the additional feature that the variances of both $\eta_i$ and
$\varepsilon_i$ have been quantified. Because these measurement errors may vary
considerably in magnitude between stars, it is beneficial to explicitly take
this between-star heterogeneity into account.

Equations~(\ref{linearRelation}) -- (\ref{drie}) yield the following relationship in terms of observable quantities:
\begin{equation}
\label{modelObservedQuantities}
S_i = \alpha + \beta \ (\varpi_i - \eta_i) + \varepsilon_{s,i} + \varepsilon_{m,i}.
\end{equation}
Here we decomposed the error of $S_i$ in a purely measurement error component,
$\varepsilon_{s,i}$ with star-specific known variance $\sigma^2_{s,i}$, and an
error component capturing variability in the relationship between $\varpi_i$ and
$S_i$, denoted by $\varepsilon_{m,i}$, for which we assume a constant but unknown
variance $\sigma^2$. Likewise, the star-specific variance of $\eta_i$ is denoted
by $\sigma^2_{p,i}$. If the relationship is indeed linear, then 
estimates of $\sigma^2$ close to zero are expected.  Furthermore, if the linear relationship
coincides with the bisector, then  $\alpha$ is expected to be close to zero and
$\beta$  close to one.
From Eq.~(\ref{modelObservedQuantities}) it follows that
\begin{align}
{\rm E}(S_i)   &= \alpha + \beta \varpi_i,   \label{ESi}  \\
{\rm Var}(S_i) &= \beta^2 \sigma^2_{p,i} + \sigma^2_{s,i} + \sigma^2,  \label{VarSi}
\end{align}
implying that the parameters can be estimated by maximising the likelihood, for example based on a normal distribution
\begin{equation}
S_i \sim N(\alpha+\beta \varpi_i,\ \beta^2\sigma^2_{p,i}+\sigma^2_{s,i}+\sigma^2)
\label{SiNormal}
\end{equation}
or the moments derived therefrom.
We note that by setting $\sigma^2_{p,i}=\sigma^2_{s,i}=0$, an ordinary linear
regression follows with homoskedastic measurement error (contrary to what
is observed).
Ordinary linear regression is expected to yield similar regression estimates as
the errors-in-variables 
model when the uncertainty in the dependent variable is considerably
larger than the uncertainty in the predictor; in the reverse case, a different
regression is obtained, an effect termed regression attenuation in statistics.

We fitted model (\ref{SiNormal}) using the SAS procedure {\tt NLMIXED}
\citep{SAS} and list the results in Tables\,\ref{table:DwarfsWithErrors} and \ref{table:DwarfsWithErrorsOtherDirection}
(including the measurement errors), and \ref{table:DwarfsWithoutErrors} and
\ref{table:DwarfsWithoutErrorsOtherDirection}
(ignoring the measurement errors).
\begin{table}
    \caption[]{Parameters of the errors-in-variables model (\ref{ESi})-(\ref{SiNormal}), 
    fitted to 22 nearby pulsators, taking the measurement errors into account.
    The values of $\alpha$ and $\sigma$ are expressed in milliarcsec, while $\beta$ is dimensionless.}
    \label{table:DwarfsWithErrors}
    \vspace*{-7mm}
    $$
    \begin{array}{cccc}
        \hline
        \noalign{\smallskip}
                 &  {\rm Estimate} & \rm{Std.~Error} & {\rm 95\%\ confidence\ interval} \\
        \noalign{\smallskip}
        \hline
        \noalign{\smallskip}
        \alpha   & 0.437  &  0.333 & [-0.254,\ 1.128] \\
        \beta    & 0.999  &  0.031 & [0.935,\ 1.062] \\
        \sigma^2 & 0.197  &  0.163 & [-0.140,\ 0.537] \\
        \noalign{\smallskip}
        \hline
    \end{array}
    $$
\end{table}
As expected from the plot, in both cases the value 0 (resp.~1) is in the 95\% confidence interval of the intercept $\alpha$ (resp.~the slope $\beta$), clearly indicating that the bisector cannot be rejected as a plausible model.

\section{{\it Kepler\/} red giants}

\begin{figure}
    \centering
    \includegraphics[width=0.5\textwidth]{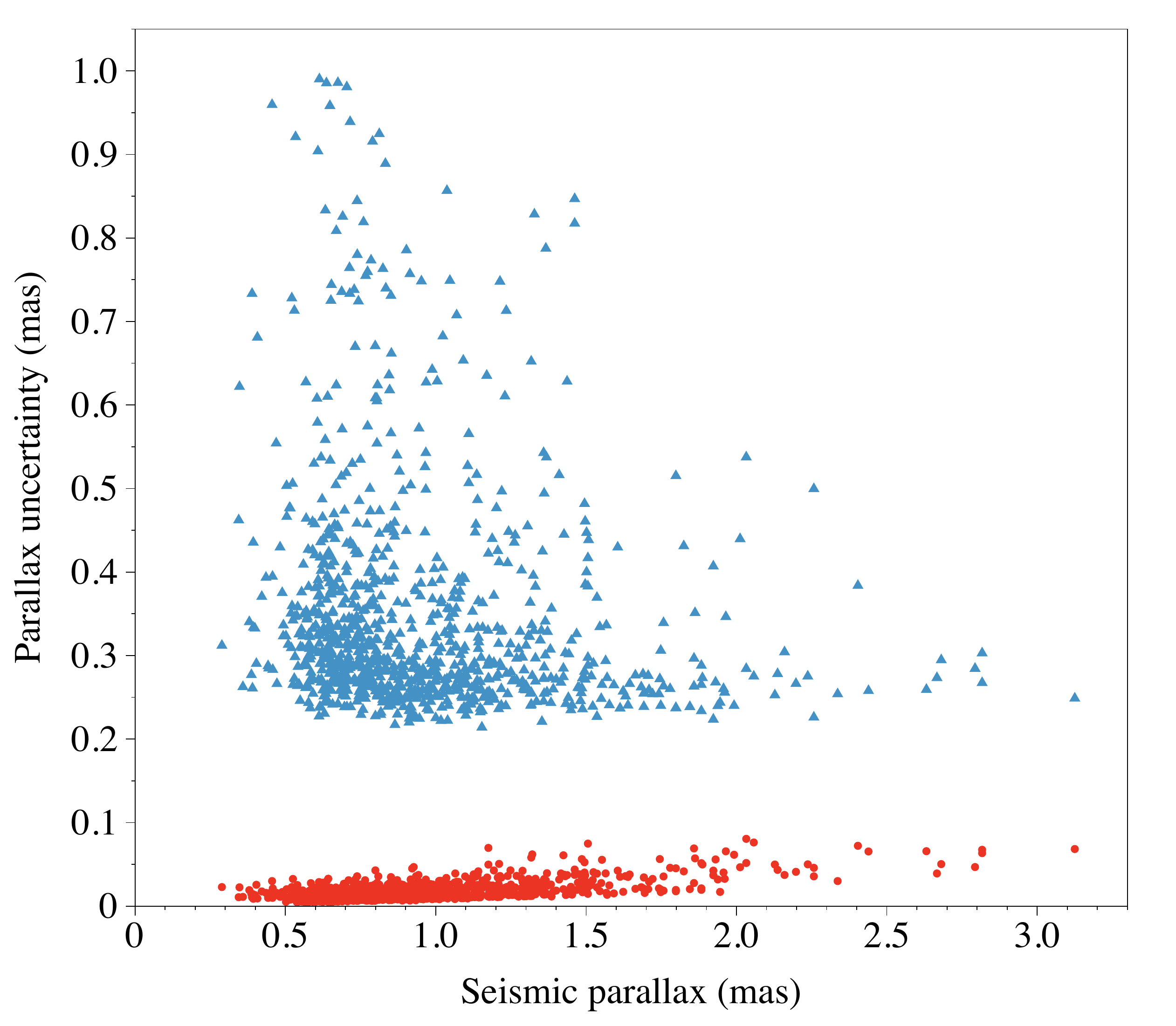}
    \caption{Uncertainties on TGAS (blue triangles) and seismic (red circles) parallaxes
             as a function of the seismic parallax of a sample of 938 red giants.}
    \label{fig:parUncertaintiesRGs}
\end{figure}

Relying on a Bayesian framework and a similar procedure for reddening
corrections to those adopted by
\cite{SilvaAguirre2012}, \citet{Rodrigues2014} derived distances to 1989 red
giants observed by {\it Kepler\/} and followed up spectroscopically with APOGEE
\citep{Pinsonneault2014}. This resulted in seismic distances between 0.5 and 5
kpc, with relative uncertainties of less than 2\%. For 938 of these giants we were
able to find a reliable crossmatch in the Gaia DR1 TGAS catalogue. We discarded stars 
with negative TGAS parallaxes as we  compare them later on with asteroseismic
parallaxes that were imposed to be positive.  The red
giants are considerably more distant than the pulsating dwarfs, leading to much more uncertain TGAS parallaxes. Figure~\ref{fig:parUncertaintiesRGs} shows that
the uncertainties of the TGAS parallaxes are substantially larger than those
obtained from the seismic distance.
Hence the current TGAS parallaxes cannot be used to calibrate the seismic values,
but a reverse validation is meaningful at this stage of the Gaia mission. We carried out
such a validation using the same methodology as in the previous section. We
prefer to analyse parallaxes rather than distances in order to avoid having to take the
reciprocal of the more variable measurements. We note that, unlike for the dwarfs, we have no information on their
possible binarity; the current TGAS parallaxes and their error estimates
assume that they are single stars. 

\begin{figure}
    \centering
    \includegraphics[width=0.5\textwidth]{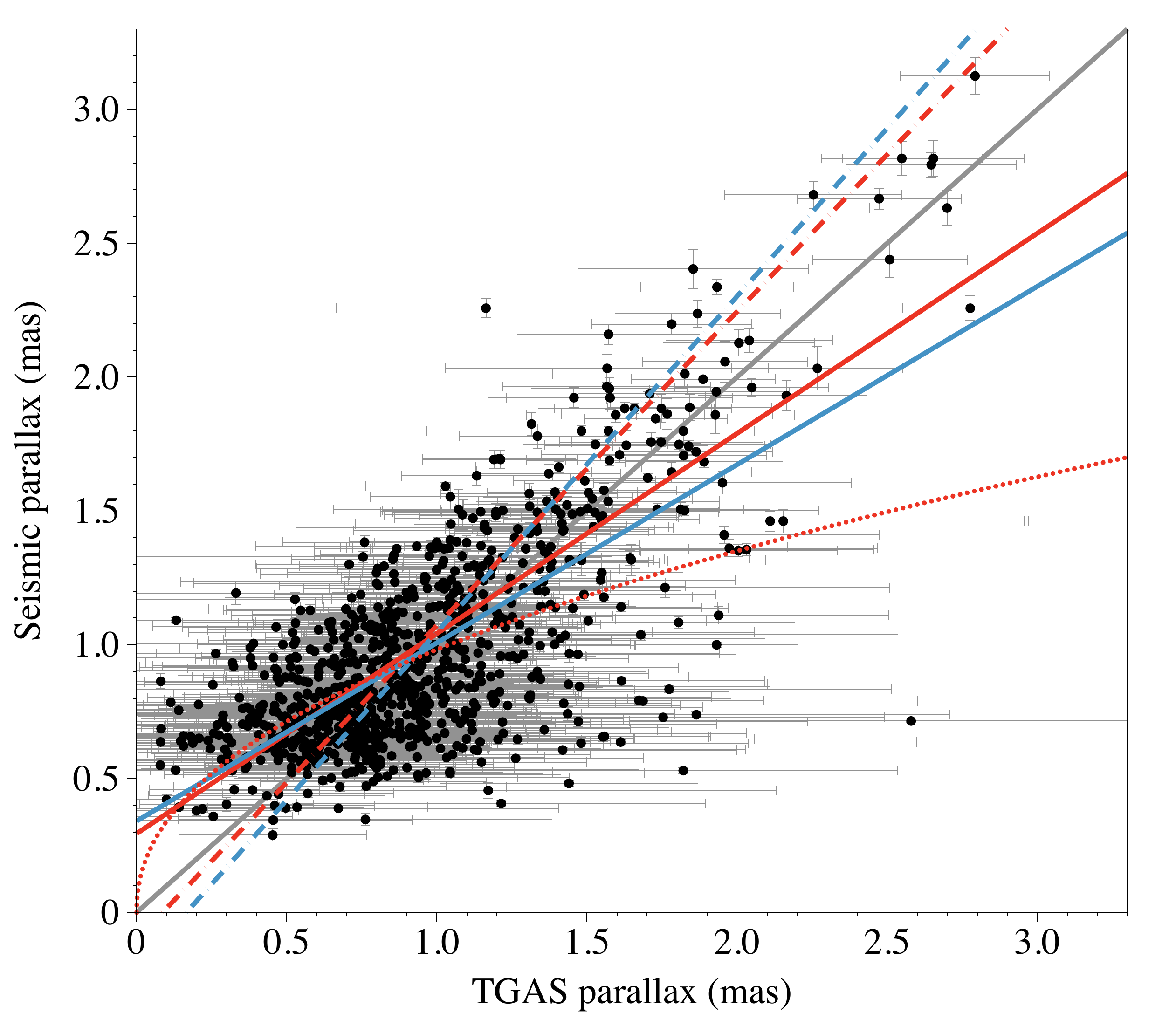}
    \caption{Seismic vs TGAS parallaxes for the sample of 938 analysed red
      giant pulsators.  The thick grey line is the bisector. Red lines: fits from an
      errors-in-variables model where the uncertainties on both parallaxes were
      taken into account; blue lines: fits obtained without taking any
      uncertainties into account. Solid lines are a fit
      using the linear model $S_i = \alpha + \beta \varpi_i$.  Dashed lines are a fit
      using the model $\varpi_i = \kappa + \rho S_i$. The curved dotted line is a fit
      using the model $\log D_i = 3+\alpha_l - (1+\beta_l) \log \varpi_i$.}
    \label{fig:parFitRGs}
\end{figure}

Figure~\ref{fig:parFitRGs} shows the astrometric and asteroseismic parallax of
the sample of 938 red giants. Clearly the relation is far less stringent than
for the dwarfs and subgiants in the previous section. This, together with the
fact that the uncertainties of one measure are much larger than  the other,
prompted us to add two other models to our analysis. First, like for
ordinary least-squares, the errors-in-variables model
(\ref{modelObservedQuantities})-(\ref{SiNormal}) is not symmetric, i.e. fitting
the seismic parallax $S_i$ as a function of the TGAS parallax $\varpi_i$ or the
other way around can lead to different results. We therefore added the
errors-in-variables model $\varpi_i = \kappa + \rho_i S_i$.

Second, the models outlined above work under the assumption that the noise sources are
additive. To assess the impact of this assumption, we add a model that assumes
multiplicative noise, which can be conveniently set up on a logarithmic scale. The relation to be tested is $10^3/D_i = \varpi_i$, where $D_i = 1/S_i$ is the asteroseismic distance expressed in pc and $\varpi_i$ the astrometric parallax expressed in mas. Using decadic logarithms, this leads to the following linear
model of the logarithmic observables $\log D_i$ and $\log \varpi_i$
\begin{equation}
\log D_i = 3+\alpha_l-(1+\beta_l)\log\varpi_i+\varepsilon_{l,i}, \label{logModel}
\end{equation}
where we assume the noise component $\varepsilon_{l,i}$ to be normally distributed,
\begin{equation}
\varepsilon_{l,i} \sim N(0,\  \sigma^2_l+\mbox{Var}(\log D_i)+(1+\beta_{l})^2 \ \mbox{Var}(\log\varpi_i)),
\end{equation}
with $\sigma_l^2$ a constant but unknown variance that captures the variability in the relationship between $\log D_i$ and $\log \varpi_i$.
The variance of the decadic logarithms of the observed quantities can be approximated as
\begin{align}
\mbox{Var}(\log D_i) &\simeq \frac{\mbox{Var}(D_i)}{\ln(10)\ D_i^2},  \label{VarLogDApprox} \\
\mbox{Var}(\log \varpi_i) & \simeq \frac{\mbox{Var}(\varpi_i)}{\ln(10)\ \varpi_i^2}. \label{VarLogParApprox}
\end{align}
The parameters were again estimated using the SAS procedure {\tt NLMIXED}.

Figure~\ref{fig:parFitRGs} synthesises  the analysis results for the fits including all  938 stars. 
The corresponding parameter estimates and their uncertainties are listed in Tables \ref{table:RedGiantsTGASvsSeismic}, and \ref{table:GiantssWithErrorsLogarithmic}--\ref{table:RedGiantsSeimicVsTGASnoErrors}. 
\begin{table}
    \caption[]{Parameters of the errors-in-variables model (\ref{ESi})-(\ref{SiNormal}).
               The fit was applied to 938 red giants, taking the measurement errors into account. The values of $\alpha$ and $\sigma$ are expressed in milliarcsec, while $\beta$ is dimensionless.}
    \label{table:RedGiantsTGASvsSeismic}
    \vspace*{-7mm}
    $$
    \begin{array}{cccc}
        \hline
        \noalign{\smallskip}
                 &  {\rm Estimate} & \rm{Std.~Error} & {\rm 95\%\ confidence\ interval} \\
        \noalign{\smallskip}
        \hline
        \noalign{\smallskip}
        \alpha   & 0.294 & 0.020  & [0.256,\ 0.333]  \\
        \beta    & 0.748 & 0.020  & [0.708,\ 0.787 ]  \\
        \sigma^2 & 0.006 & 0.004  & [-0.002,\ 0.013] \\
        \noalign{\smallskip}
        \hline
    \end{array}
    $$
\end{table}
Not surprisingly, of all linear fits, the ordinary least-squares fit $S_i = \alpha
+ \beta\varpi_i$ (solid blue line) that does not incorporate any uncertainties
deviates most from the bisector. When we do include the uncertainties of both
measures, the resulting fit (solid red line) considerably improves, but
with a slope of 0.75 it is still significantly tilted with respect to~the
bisector. This occurs in part because  the fit tries to accommodate the fact that
the asteroseismic analysis locates remote giants systematically nearer than
TGAS. We note that the intercept of the solid red line (as listed in Table
\ref{table:RedGiantsTGASvsSeismic}) is $0.29\pm0.02$ mas, which is similar to the
value of $0.3$ mas that \citet{Lindegren2016} quote as the typical systematic
error on the parallax, depending on position and colour. It is unclear whether
this a coincidence or not.

Since the uncertainties on the TGAS parallaxes are so much larger than the seismic values, the fit of the inverse model $\varpi_i = \kappa + \rho S_i$ treats the seismic parallaxes almost as fixed and known, similar to ordinary weighted least-squares. Also in this case, taking the uncertainties into account (dashed red line) tilts the fit closer to the bisector than ignoring them, although the difference is less pronounced than for the first model.

We also verified whether model (\ref{logModel}) with multiplicative noise would be appropriate
for this dataset. The resulting fit is shown as a red dotted line in Fig.~\ref{fig:parFitRGs},
and the corresponding parameter estimates are listed in Table \ref{table:GiantssWithErrorsLogarithmic}.

The result of each of these models is that the 95\% confidence interval for the
slope never contains 1, and the corresponding interval for the intercept never
contains 0. That is, the bisector is not a plausible model.

\section{Conclusions}

Astrometric parallaxes in the Gaia DR1 TGAS catalogue have been compared with
published asteroseismic distances of pulsating dwarfs and giants using an
errors-in-variables approach.  Deviation from the bisector would imply that
either the models of stellar interiors combined with reddening models at the basis of 
the asteroseismic distances need revision, or that there is an unknown 
systematic uncertainty on the current version of the astrometric parallaxes, 
or both.

Proper statististical analysis of the two parallax estimates is {a priori\/}
more complicated than ordinary least-squares as we now have uncertainties on both
estimates of one and the same quantity.  We therefore set up an
errors-in-variables model,  on a linear scale and on a logarithmic scale, to
do the fitting. For the 22 dwarfs and subgiants of \citet{SilvaAguirre2012}, the
results reveal excellent agreement between the two distances,
re-confirming the asteroseismic achievement for these stars now that we have
more precise parallaxes from TGAS. 

For our sample of 938 giants taken from \citet{Rodrigues2014} and crossmatched
with TGAS, the relative uncertainties on the astrometric parallaxes are much larger than
on the seismic distances, turning the latter into a valuable instrument to
validate the former. All the models we applied -- an errors-in-variables model,
ordinary least-squares, and a logarithmic model -- lead to a significant
difference between astrometric and asteroseismic parallaxes. There can be
several underlying causes. The uncertainties of the TGAS parallaxes
may be underestimated or could be subject to systematic errors. On the other
hand, 
interstellar extinction corrections and/or too poorly known bulk metallicity may have introduced 
a systematic uncertainty for the asteroseismic parallax. Given that all stars in the 
Gaia DR1 TGAS catalogue were assumed to be single, binarity may also be part of the
cause. 

Expectations are that the accuracy of the Gaia astrometric distance measurements will surpass the seismic measurements by the end of the mission. Gaia Data Release 2 (end 2017)
will contain the astrometry of a billion stars, while Gaia Data Release 3 
(end 2018) will deliver the orbital
astrometric solutions for binaries with periods longer than 2 months. This will
allow us to improve our current research and transfer it into a quantitative
calibration of asteroseismic distances for a variety of stellar populations.

\begin{acknowledgements}
JDR and CA gratefully acknowledge the support from the Belgian Federal Science Policy Office (Belspo, Gaia-DPAC)
and from the European Research Council (ERC) under the European Union’s Horizon 
2020 research and innovation programme (grant agreement N$^\circ$670519: MAMSIE). GM gratefully acknowledges
support from IAP research Network P7/06 of the Belgian Government (Belgian Science Policy). 
This work has made use of data from the European Space Agency (ESA)
mission {\it Gaia} (\url{http://www.cosmos.esa.int/gaia}), processed by
the {\it Gaia} Data Processing and Analysis Consortium (DPAC,
\url{http://www.cosmos.esa.int/web/gaia/dpac/consortium}). Funding
for the DPAC has been provided by national institutions, in particular
the institutions participating in the {\it Gaia} Multilateral Agreement.
This research made use of the SIMBAD database and the VizieR catalogue access tool, operated at CDS, Strasbourg, 
France, and of the SAO/NASA Astrophysics Data System.
\end{acknowledgements}

\bibliographystyle{aa}
\bibliography{paper_final}

\begin{appendix}
\section{Additional parameter estimates}

In this appendix, we provide the outcome of parameter estimates using the models discussed
in the main text. Tables \ref{table:DwarfsWithoutErrors}--\ref{table:DwarfsWithErrorsLogarithmic}
concern the sample of 22 nearby dwarfs and subgiants. 
Tables \ref{table:GiantssWithErrorsLogarithmic}--\ref{table:RedGiantsSeimicVsTGASnoErrors} relate 
to the sample of 938 red giants. Tables \ref{table:DwarfsWithoutErrorsOtherDirection} and
\ref{table:DwarfsWithErrorsOtherDirection} provide the estimates of the intercept and slope for the 
reverse model $\varpi_i = \kappa + \rho S_i$, which was introduced specifically for the giants, 
but which we also fitted for the dwarfs for the sake of completeness. The fits lead to exactly
the same conclusion as the normal model (\ref{SiNormal}). Also here the value zero is well within
the 95\% confidence interval of the intercept, and the value 1 is within the corresponding interval
of the slope, indicating that the bisector is a plausible model for the relation between astrometric
and asteroseismic parallaxes. Not taking the uncertainties into account leads to an estimate of 
$\sigma^2$ (the error component capturing variability in the linear relationship), which is significantly
different from zero, but this significance disappears when the uncertainties are taken into account
showing that the linear model is adequate.
\begin{table}
\caption[]{Parameters of the model (\ref{ESi})-(\ref{SiNormal})
    fitted to 22 nearby pulsators, {without} taking the measurement errors into account. The values of
    $\alpha$ and $\sigma$ are expressed in milliarcsec, while $\beta$ is dimensionless.}
    \label{table:DwarfsWithoutErrors}
    \vspace*{-7mm}
    $$
    \begin{array}{cccc}
        \hline
        \noalign{\smallskip}
                 &  {\rm Estimate} & \rm{Std.~Error} & {\rm 95\%\ confidence\ interval} \\
        \noalign{\smallskip}
        \hline
        \noalign{\smallskip}
        \alpha   & 0.549 & 0.296  & [-0.064,\ 1.162] \\
        \beta    & 0.990 & 0.022  & [0.945,\  1.034] \\
        \sigma^2 & 0.556 & 0.168  & [0.208,\  0.904] \\
        \noalign{\smallskip}
        \hline
    \end{array}
    $$
\end{table}
\begin{table}[h!]
  \caption[]{Parameters of the model
    (\ref{ESi})-(\ref{SiNormal}) using the relation $\varpi_i = \kappa + \rho S_i$, applied
    to 22 nearby pulsators, without taking the measurement errors into account. The values of  
    $\kappa$ and $\sigma$ are expressed in milliarcsec, while $\rho$ is dimensionless.}
    \label{table:DwarfsWithoutErrorsOtherDirection}
    \vspace*{-7mm}
    $$
    \begin{array}{cccc}
        \hline
        \noalign{\smallskip}
                 &  {\rm Estimate} & \rm{Std.~Error} & {\rm 95\%\ confidence\ interval} \\
        \noalign{\smallskip}
        \hline
        \noalign{\smallskip}
        \kappa   & -0.430 & 0.306  & [-1.065,\ 0.205] \\
        \rho     &  1.000 & 0.022  & [0.955 ,\ 1.045] \\
        \sigma^2 &  0.562 & 0.170  & [0.211,\  0.914] \\
        \noalign{\smallskip}
        \hline
    \end{array}
    $$
\end{table}
\begin{table}[h!]
  \caption[]{Parameters of the model
    (\ref{ESi})-(\ref{SiNormal}) using the relation $\varpi_i = \kappa + \rho S_i$. 
    The fit was applied to 22 nearby pulsators, taking the measurement errors into account. The values of
    $\kappa$ and $\sigma$ are expressed in milliarcsec, while $\rho$ is dimensionless.}
    \label{table:DwarfsWithErrorsOtherDirection}
    \vspace*{-7mm}
    $$
    \begin{array}{cccc}
        \hline
        \noalign{\smallskip}
                 &  {\rm Estimate} & \rm{Std.~Error} & {\rm 95\%\ confidence\ interval} \\
        \noalign{\smallskip}
        \hline
        \noalign{\smallskip}
        \kappa   & -0.240 & 0.327 & [-0.917,\ 0.437] \\
        \rho     &  0.982 & 0.029 & [0.921 ,\ 1.042] \\
        \sigma^2 &  0.185 & 0.158 & [-0.142,\ 0.511] \\
        \noalign{\smallskip}
        \hline
    \end{array}
    $$
\end{table}
\begin{table}[h!]
    \caption[]{Parameters of the model (\ref{logModel}) on the logarithmic scale, 
    fitted to 22 nearby pulsators, taking the measurement errors into account.}
    \label{table:DwarfsWithErrorsLogarithmic}
    \vspace*{-7mm}
    $$
    \begin{array}{cccc}
        \hline
        \noalign{\smallskip}
                 &  {\rm Estimate} & \rm{Std.~Error} & {\rm 95\%\ confidence\ interval} \\
        \noalign{\smallskip}
        \hline
        \noalign{\smallskip}
        \alpha_l   &  -0.054 & 0.032  & [-0.119,\   0.012] \\
        \beta_l    &  -0.034 & 0.029  & [-0.095,\   0.027] \\
        \sigma_l^2 &   0.00038 & 0.00028 & [-0.00020,\ 0.00096] \\
        \noalign{\smallskip}
        \hline
    \end{array}
    $$
\end{table}
\begin{table}[h!]
    \caption[]{Parameters of the model (\ref{logModel}) on the logarithmic scale, 
    fitted to 938 pulsating red giants, taking the measurement errors into account.}
    \label{table:GiantssWithErrorsLogarithmic}
    \vspace*{-7mm}
    $$
    \begin{array}{cccc}
        \hline
        \noalign{\smallskip}
                 &  {\rm Estimate} & \rm{Std.~Error} & {\rm 95\%\ confidence\ interval} \\
        \noalign{\smallskip}
        \hline
        \noalign{\smallskip}
        \alpha_l   &  0.008 & 0.005  & [-0.001,\   0.018] \\
        \beta_l    & -0.540 & 0.028  & [-0.596,\  -0.485] \\
        \sigma_l^2 & -0.174 & 0.0005 & [-0.175,\  -0.173] \\
        \noalign{\smallskip}
        \hline
    \end{array}
    $$
\end{table}
\begin{table}
    \caption[]{Parameters of the model  
               (\ref{ESi})-(\ref{SiNormal}) using the relation $\varpi_i = \kappa + \rho S_i$, applied to a sample of 938 red giants,
               taking the measurement errors into account. The values of  $\kappa$ and $\sigma$ are expressed in milliarcsec, while $\rho$ is dimensionless.}
    \label{table:RedGiantsSeismicVsTGAS}
    \vspace*{-7mm}
    $$
    \begin{array}{cccc}
        \hline
        \noalign{\smallskip}
                 &  {\rm Estimate} & \rm{Std.~Error} & {\rm 95\%\ confidence\ interval} \\
        \noalign{\smallskip}
        \hline
        \noalign{\smallskip}
        \kappa   & 0.086  & 0.022  & [0.042,\ 0.130] \\
        \rho     & 0.853  & 0.020  & [0.813,\ 0.892] \\
        \sigma^2 & -0.024 & 0.002  & [-0.029,\ -0.019] \\
        \noalign{\smallskip}
        \hline
    \end{array}
    $$
\end{table}
\begin{table}
    \caption[]{Parameters of the model (\ref{ESi})-(\ref{SiNormal}).
               The fit was applied to 938 red giants, {without} taking the measurement errors into account. The values of  $\alpha$ and $\sigma$ are expressed in milliarcsec, while $\beta$ is dimensionless.}
    \label{table:RedGiantsTGASvsSeismicNoErrors}
    \vspace*{-7mm}
    $$
    \begin{array}{cccc}
        \hline
        \noalign{\smallskip}
                 &  {\rm Estimate} & \rm{Std.~Error} & {\rm 95\%\ confidence\ interval} \\
        \noalign{\smallskip}
        \hline
        \noalign{\smallskip}
        \alpha   & 0.341 & 0.021 & [0.300,\ 0.383]  \\
        \beta    & 0.666 & 0.020 & [0.626,\ 0.706]  \\
        \sigma^2 & 0.072 & 0.003 & [0.066,\ 0.079]  \\
        \noalign{\smallskip}
        \hline
    \end{array}
    $$
\end{table}
\begin{table}
    \caption[]{Parameters of the model
               (\ref{ESi})-(\ref{SiNormal}) using the relation $\varpi_i = \kappa + \rho S_i$, applied to a sample of 938 red giants, {without} taking the measurement errors into account. The values of  $\kappa$ and $\sigma$ are expressed in milliarcsec, while $\rho$ is dimensionless.}
    \label{table:RedGiantsSeimicVsTGASnoErrors}
    \vspace*{-7mm}
    $$
    \begin{array}{cccc}
        \hline
        \noalign{\smallskip}
                 &  {\rm Estimate} & \rm{Std.~Error} & {\rm 95\%\ confidence\ interval} \\
        \noalign{\smallskip}
        \hline
        \noalign{\smallskip}
        \kappa   & 0.166 & 0.025 & [0.116,\ 0.216] \\
        \rho     & 0.797 & 0.024 & [0.749,\ 0.845] \\
        \sigma^2 & 0.086 & 0.004 & [0.079,\ 0.094] \\
        \noalign{\smallskip}
        \hline
    \end{array}
    $$
\end{table}
\end{appendix}
\end{document}